\documentclass[12pt]{article}
\usepackage{amsmath}
\usepackage{graphicx}
\usepackage{color}
\begin{document}
\baselineskip=18 pt
\begin{center}
{\large{\bf  Axial symmetry cosmological constant vacuum solution of field equations with a curvature singularity, closed time-like curves and deviation of geodesics }}
\end{center}

\vspace{.5cm}

\begin{center}
{\bf Faizuddin Ahmed}\footnote{\bf faizuddinahmed15@gmail.com (Corresponding author)}\\
{\bf Maryam Ajmal Women's College of Science \& Technology, Hojai-782435, Assam, India} \\ \vspace{0.2cm}
{\bf Bidyut Bikash Hazarika}\footnote{\bf bidyutbikash116@gmail.com} \mbox{and} 
{\bf Debojit Sarma}\footnote{\bf sarma.debojit@gmail.com}\\
{\bf Department of Physics, Cotton University, Guwahati-781001, Assam, India}
\end{center}

\vspace{.5cm}

\begin{abstract}

In this paper, we present a type D, non-vanishing cosmological constant, vacuum solution of the Einstein's field equations, extension of an axially symmetric, asymptotically flat vacuum metric with a curvature singularity. The space-time admits closed time-like curves (CTCs) that appear after a certain instant of time from an initial spacelike hypersurface, indicating it represents a time-machine space-time. We wish to discuss the physical properties and show that this solution can be interpreted as gravitational waves of Coulomb-type propagate on anti-de Sitter space backgrounds. Our treatment focuses on the analysis of the equation of geodesic deviations.

\end{abstract}



\vspace{.5cm}

\section{Introduction}

Closed time-like curves constitute one of the most intriguing aspects of general relativity. The first solution of the field equations admitting closed time-like curves (CTCs) is the G\"{o}del rotating Universe \cite{KG}. It represents a rotating universe and is axially symmetric, given by
\begin{equation}
ds^2=dr^2+dz^2+(\sinh^2 r-\sinh^4 r)\,d\theta^2+2\,\sqrt{2}\,\sinh^2\,r\,d\theta\,dt-dt^2
\label{godel}
\end{equation}
The coordinates are in the ranges $0 \leq r < \infty$, $-\infty < z < \infty$, $-\infty < t < \infty$ and $\theta$ is periodic. For some $r > r_0$, the metric function, $g_{\theta\theta}=\sinh^2 r-\sinh^4 r$ becomes negative. The circle defined by $r > r_0$, and $t=0=z$ will be time-like everywhere. This condition is fulfilled when $r > r_0=\ln(1+\sqrt{2})$ which is the condition for existence of CTCs in the G\"{o}del space-time because one of the coordinates $\theta \in [0, 2\pi ]$ is periodic. Next one is the van Stockum space-time \cite{Sto}, which pre-dates the G\"{o}del solution, and was shown later to have CTCs \cite{Tip}. Examples of space-time admitting CTCs including NUT-Taub metric \cite{NUT,NUT2,NUT3,NUT4}, Kerr and Kerr-Newmann black holes solution \cite{Kerr,Cart,AT}, Gott time-machine \cite{Gott}, Grant space-time \cite{Grant}, Krasnikov tube \cite{Kras}, Bonnor's metrics \cite{Bon1,Bon2,Bon3,Bon4,Coll,Bon5}, and others \cite{Gurs,Rosa,Evs,GERG,EPJP,AOP,PTEP2,CTP,CTP2,PTEP3,AOP2,JPCO,TMP,EPJC,PTEP,GC,GC2}. Space-time with causality violating curves are classified as either eternal or true time-machine space-times. In Eternal time machine space-time case, CTCs always pre-exist. In this category would be \cite{KG} or \cite{Sto} (see also, Refs. \cite{GERG,AOP,CTP,TMP,GC2}). A true time machine space-time is the one in which CTCs evolve at a particular instant of time from an initial spacelike hypersurface in a causally well-behaved manner satisfying all the energy conditions with known type of matter fields. In this category, the Ori time machine space-time \cite{Ori} is considered to be most remarkable. But the matter sources satisfying all the energy conditions is of unknown type in this space-time. Most of the time machine models suffer from one or more drawbacks. For space-time admitting CTCs, the matter-energy sources must be realistic, that is, the stress-energy tensor must be of known type of matter fields which satisfy all the energy conditions. Many space-time model, for examples, traversable wormholes \cite{MTY1,MTY2}, warp drive models \cite{Alcu,Ever,Ever2,Lobo} violate the Weak energy condition (WEC), which states that $T_{\mu\nu}\,U^{\mu}\,U^{\nu}\geq 0$ for a time-like tangent vector field $U^{\mu}$, that is, the energy-density must be non-negative. Some other space-times admitting CTCs violate the strong energy condition (SEC) ({\it e. g.}, Refs. \cite{Ori22,Ori33,Ori44,KD}), which states that $(T_{\mu\nu}-\frac{1}{2}\,g_{\mu\nu}\,T)\,U^{\mu}\,U^{\nu}\geq 0$. In addition, some solutions does not admit a partial Cauchy surface (initial spacelike hypersurface) ({\it e. g.}, Refs. \cite{KG,Mall}) and/or CTCs come from infinity ({\it e. g.}, Refs. \cite{Gott,Grant}). In addition, there is a curvature singularity in some solutions admitting CTCs \cite{Tip,GC,Mall,Mall2,Ori2,AHEP,AHEP2,AHEP3}.

In literature, only handful solutions of the Einstein's field equations with the stress energy tensor in \cite{KG,EPJC,PTEP} and type N Einstein space-time in \cite{AOP2} have a negative cosmological constant. In this work, we try to construct a type D Einstein space-time with a negative cosmological constant which wasn't study earlier. The cosmological constant plays a vital role in explaining the dynamics of the universe. A tiny positive cosmological constant neatly explains the late time accelerated expansion of the universe. Indeed our universe is observed to be undergoing a de Sitter (dS) type expansion in the present epoch. For a negative cosmological constant, space-time is labelled as anti-de Sitter (AdS) space. The AdS space has been a subject of intense study in recent times on account of the celebrated AdS/CFT correspondence \cite{JMM} which provides a link between a quantum theory of gravity on an asymptotically AdS space and a lower dimensional conformal field theory (CFT) on its boundary.

\section{Review of a type D vacuum space-time with a curvature singularity and CTCs \cite{AHEP} }

In Ref. \cite{AHEP}, a type D axially symmetric, asymptotically flat vacuum solution of the field equations with zero cosmological constant, was constructed. This vacuum metric is as follow
\begin{eqnarray}
ds^2&=&-\cosh t\,\mbox{coth} t\,\sinh^2 r\,dt^2+\cosh^2 r\,\sinh r\,dr^2+\mbox{csch} r\,dz^2\nonumber\\
&+&\sinh^2 r\,(2\,\sqrt{2}\,\cosh t\,dt\,d\phi-\sinh t\,d\phi^2).
\label{debojit}
\end{eqnarray}
After doing a number of transformations into the above metric, we arrive at the following
\begin{equation}
ds^2=\cosh^2 r\,\sinh r\,dr^2+\mbox{csch} r\,dz^2-\sinh^2 r\,(2\,dt\,d\phi+t\,d\phi^2).
\label{debojit2}
\end{equation}
The Kretschmann scalar of the above metric is
\begin{eqnarray}
K=R_{\mu\nu\rho\sigma}\,R^{\mu\nu\rho\sigma}=\frac{12}{\sinh^{6} r}.
\label{krets}
\end{eqnarray}
For constant $r,z$, the metric (\ref{debojit2}) reduces to conformal Misner metric in 2D 
\begin{equation}
ds^2=\Omega\,(-2\,dt\,d\phi-t\,d\phi^2),
\label{1}
\end{equation}
where $\Omega=\sinh^2 r$ is the conformal factor.

In the context of CTCs, the Misner space metric in 2D is interesting because CTCs appear after a certain instant of time from causally well-behaved conditions. The metric for the Misner space in 2D \cite{Mis} is given by 
\begin{equation}
ds_{Mis}^{2}=-\,2\,dT\,dX-T\,dX^2
\label{mis}
\end{equation}
where $-\infty < T < \infty$ but the co-ordinate $X$ is periodic locally. The metric (\ref{mis}) is regular everywhere as $det\,g=-1$ including at $T=0$. The curves $T=T_0$, where $T_0$ is a constant, are closed since $X$ is periodic. The curves $T < 0$ are spacelike, $T>0$ are time-like, while the null curves $T=0$ form the chronology horizon. The second type of curves, namely, $T=T_0 > 0$ are closed time-like curves. Therefore, the metric (\ref{debojit}) or (\ref{debojit2}) is a four-dimensional generalization of Misner space in curved space-time. Note that the above space-time is vacuum solution of field equations, a Ricci flat, that is, $R_{\mu\nu}=0$. Li \cite{Li} constructed a Misner-like AdS space-time, a time-machine model. Levanony {\it et al.} \cite{Leva} constructed a three-, four-dimensional generalization of flat Misner space metric.

In this paper, we extend the above Ricci flat space-time (\ref{debojit}) to the Einstein space-times of Petrov type D, which satisfy the following conditions
\begin{equation}
R_{\mu\nu}=\Lambda\,g_{\mu\nu},\quad R=4\,\Lambda,\quad \Lambda <0 \quad \mbox{or} \quad \Lambda >0
\label{condition}
\end{equation}
It is an anti-de Sitter-like space if $\Lambda <0$ and de Sitter like if $\Lambda>0$. The extended space-time satisfies all the basic requirements (see details in Ref. \cite{AOP2}) for a time machine space-time except one, that is, this new model is not free from curvature singularity.

\section{Analysis of a cosmological constant vacuum space-time} 
  
Consider the following line element, a modification of the metric (\ref{debojit}) given by
\begin{eqnarray}
ds^2&=&\sinh^2 r\,(-\cosh t\,\mbox{coth} t\,dt^2+2\sqrt{2}\,\cosh t\,dt\,d\phi-\sinh t\,d\phi^2)\nonumber\\
&+&\frac{dr^2}{(\alpha\,\mbox{csch} r\,\mbox{sech}^2 r+\beta^2\,\mbox{tanh}^2 r)} +(\alpha\,\mbox{csch} r + \beta^2\,\sinh^2 r)\,dz^2.
\label{temp}
\end{eqnarray}
Here $\alpha$ is a positive constant and $\beta$ is real. The coordinates are labelled $x^0=t$, $x^1=r$, $x^2=\phi$, and $x^3=z$. The ranges of the coordinates are 
\begin{equation}
-\infty < t < \infty,\quad,\quad 0 \leq r < \infty,\quad -\infty < z < \infty
\end{equation}
 and $\phi$ is a periodic coordinate $\phi\sim\phi+\phi_0$, with $\phi_0>0$. The metric is Lorentzian with signature $(-,+,+,+)$ and the determinant of the corresponding metric tensor $g_{\mu\nu}$,
\begin{equation}
det\;g=-\cosh^2 r\,\sinh^4 r\,\cosh^2 t. 
\end{equation}

Now we have evaluated the Ricci tensor $R_{\mu\nu}$ of the space-time (\ref{temp}) as follows:
\begin{eqnarray}
&&R_{tt}=3\,\beta^2\,\cosh t\,\mbox{coth} t\,\sinh^2 r,\nonumber\\
&&R_{t\phi}=R_{\phi t}=-3\,\sqrt{2}\,\beta^2\,\cosh t\,\sinh^2 r,\nonumber\\
&&R_{rr}=-3\,\beta^2\,\left (\frac{\cosh^2 r}{\alpha\,\mbox{csch} r+\beta^2\,\sinh^2 r} \right),\nonumber\\
&&R_{\phi\phi}=3\,\beta^2\,\sinh^2 r\,\sinh t,\nonumber\\
&&R_{zz}=-3\,\beta^2\,\left (\alpha\,\mbox{csch} r+\beta^2\,\sinh^2 r \right).
\label{ricci-components}
\end{eqnarray}
The Ricci scalar is given by 
\begin{equation}
R^{\mu}_{\,\mu}=R=-12\,\beta^2.
\label{scalar}
\end{equation}
Using the metric tensor components of the above space-time, we have found that the Ricci tensor
\begin{equation}
R_{\mu\nu}=-3\,\beta^2\,g_{\mu\nu} \quad (\mu,\nu=0,1,2,3).
\label{scalar2}
\end{equation}
And the Einstein tensor $G_{\mu\nu}$ are
\begin{equation}
G^{\mu}_{\,\nu}=3\,\beta^2\,\mbox{diag}\,(1,1,1,1).
\label{scalar3}
\end{equation}
From the Einstein's field equations $G_{\mu\nu}+\Lambda\,g_{\mu\nu}=0$ and from eq. (\ref{scalar3}), we have
\begin{equation}
3\,\beta^2=-\Lambda \Rightarrow \beta=\pm\,\sqrt{-\frac{\Lambda}{3}}\quad,\quad \Lambda < 0.
\end{equation}
Thus from above analysis it is clear that the space-time considered by (\ref{temp}) is an example of the class of Einstein space of anti-de Sitter-type and satisfies eq. (\ref{condition}) for a negative cosmological constant. We have shown later that the space-time possesses a curvature singularity at $r\rightarrow 0$.

An interesting property of the metrics (\ref{temp}) is that it reduces to 2D Misner space metric \cite{Mis} for constant $r, z$. For that, we do the following transformations  
\begin{equation}
t \rightarrow \mbox{sinh}^{-1} t\quad,\quad \phi \rightarrow \phi+(\sqrt{2}+1)\, \mbox{ln} t
\label{trans}
\end{equation}
into the metric (\ref{temp}) (replacing $\beta^2 \rightarrow -\frac{\Lambda}{3}$), we arrive at the following line element
\begin{eqnarray}
ds^2&=&\sinh^2 r\,(-2\,dt\,d\phi - t\,d\phi^2)+\frac{\cosh^2 r\,dr^2}{\left (\alpha\,\mbox{csch}r - \frac{\Lambda}{3}\,\sinh^2 r \right)}\nonumber\\
&+&\left (\alpha\,\mbox{csch} r-\frac{\Lambda}{3}\,\sinh^2 r \right)\,dz^2.
\label{temp2}
\end{eqnarray}

For constant $r= r_0>0$ and $z=z_0$, the metric (\ref{temp2}) becomes 
\begin{equation}
ds^2_{conf}=\sinh^2 r\,(-2\,dt\,d\phi-t\,d\phi^2)=\Omega\,ds^2_{Mis},
\label{line}
\end{equation}
a conformal Misner space metric in 2D where, $\Omega$ is the conformal factor. Therefore, the space-time admits CTC for $t=t_0>0$ similar to the Misner space discussed earlier.

We check whether the CTCs evolve from an initially spacelike $t=constant$ hypersurface (and thus $t$ is a time coordinate). This is determined by calculating the norm of the vector $\nabla_\mu t$ \cite{Ori} (or alternately from the value of $g^{tt}$ in the inverse metric tensor $g^{\mu\nu}$). A hypersurface $t=constant$ is spacelike when $g^{tt}<0$ at $t<0$, time-like when $g^{tt}>0$ for $t>0$ and null $g^{tt}=0$ for $t=0$. For the metric (\ref{temp}), we have
\begin{equation}
\nabla_\mu t\nabla^\mu t=g^{tt}=\frac{\sinh t}{\sinh^2 r\,\cosh^2 t}\quad
\label{hype}
\end{equation}
Thus a hypersurface $t=constant$ is spacelike for $t<0$, time-like for $t>0$ and null at $t=0$. We restrict our analysis to $r>0$ otherwise no CTCs will be formed. Thus the spacelike $t=constant< 0$ hypersurface can be chosen as initial hypersurface over which initial data may be specified. There is a Cauchy horizon at $t=t_0=0$ called the chronology horizon, which separates the causal past and future in a past directed and future directed manner. Hence the space-time evolves from a partial Cauchy surface ({\it i. e.} initial spacelike hypersurface) in a causally well-behaved, up to a moment, {\it i. e.,} a null hypersurface $t=t_0=0$ and the formation of CTCs takes place from causally well-behaved initial conditions. The evolution of CTC is thus identical to the case of the Misner space. 

That the space-time represented by (\ref{temp}) satisfy the requirements of axial symmetry is clear from the following. Consider the Killing vector ${\bf{\eta}}=\partial_{\phi}$ having the normal form
\begin{equation}
\eta^{\mu}=\left(0,0,1,0 \right ).
\label{killi}
\end{equation}
Its co-vector form  
\begin{equation}
\eta_{\mu}=\sinh^2 r\left(\sqrt{2}\,\cosh t,0,-\sinh t,0\right ).
\label{killi1} 
\end{equation}
The vector (\ref{killi1}) satisfies the Killing equation $\eta_{\mu\,;\,\nu}+\eta_{\nu\,;\,\mu}=0$. The space-time is axial symmetry if the norm of the Killing vector $\eta^{\mu}$ vanish on the axis {\it i. e.} at $r=0$ (see \cite{Mars1,Mars2} and references there in). In our case 
\begin{equation}
X=|\eta_{\mu}\,\eta^{\mu}|=|g_{\phi\phi}|=|-\sinh t\,\sinh^2 r|\rightarrow 0,
\label{killi2}
\end{equation}
as $r \rightarrow 0$.

The metric has a curvature singularity at $r=0$. We find that the Kretschmann scalar is
\begin{equation}
K=R_{\mu\nu\rho\sigma}\,R^{\mu\nu\rho\sigma}=\frac{8\,\Lambda^2}{3}+\frac{12\,\alpha^2}{\sinh^6 r}\quad
\label{kret}
\end{equation}
We can see that the scalar curvature diverge at $r \rightarrow 0$ which indicates the space-time possesses a curvature singularity. In addition, the Krestchmann scalar becomes $K \rightarrow \frac{8\,\Lambda^2}{3}$ for $r \rightarrow \infty$ indicating that the metric (\ref{temp}) is asymptotically anti-de Sitter-like space radially \cite{AW}.

\subsection{Classification and physical Interpretation of the space-times}

Here we first classify the space-time according to the Petrov classification scheme, and then analyze the effect of local fields of the solution. We construct a set of null tetrad $(\bf {k, l, m, \bar{m}})$ \cite{HS} for the space-time (\ref{temp}). Explicitly these co-vectors are 
\begin{eqnarray}
\label{x1}
&&k_{\mu}=\frac{\sinh r}{\sqrt{2}}\,\left (\frac{\cosh t}{\sqrt{\sinh t}},0,(-\sqrt{2}+1)\,\sqrt{\sinh t},0 \right),\\
\label{x2}
&&l_{\mu}=\frac{\sinh r}{\sqrt{2}}\,\left (\frac{\cosh t}{\sqrt{\sinh t}},0,-(\sqrt{2}+1)\,\sqrt{\sinh t},0 \right),\\
\label{x3} 
&&m_{\mu}=\frac{1}{\sqrt{2}}\,\left (0,\frac{\cosh r}{\sqrt{\alpha\,\mbox{csch} r-\frac{\Lambda}{3}\,\sinh^2 r}},0,i\,\sqrt{\alpha\,\mbox{csch} r-\frac{\Lambda}{3}\,\sinh^2 r} \right),\\
\label{x4}
&&\bar{m}_{\mu}=\frac{1}{\sqrt{2}}\,\left (0,\frac{\cosh r}{\sqrt{\alpha\,\mbox{csch} r- \frac{\Lambda}{3}\,\sinh^2 r}},0,-i\,\sqrt{\alpha\,\mbox{csch} r-\frac{\Lambda}{3}\,\sinh^2 r} \right).
\end{eqnarray}
The set of null tetrad above is such that the metric tensor for the line element (\ref{temp}) can be expressed as
\begin{equation}
g_{\mu \nu}=-k_{\mu}\,l_{\nu}-l_{\mu}\,k_{\nu}+m_{\mu}\,\bar{m}_{\nu}+\bar{m}_{\mu}\,m_{\nu} \quad 
\label{n2}
\end{equation}
The vectors (\ref{x1}), (\ref{x2}), (\ref{x3}) and (\ref{x4}) are null vector and orthogonal, except for $k_{\mu}l^{\mu}=-1$ and $m_{\mu}{\bar m}^{\mu}=1$. 

We calculate the five Weyl scalars, of these only
\begin{equation}
\Psi_2=C_{\mu\nu\rho\sigma}\,k^{\mu}\,m^{\nu}\,{\bar m}^{\rho}\,l^{\sigma}=\frac{\alpha}{2\,\sinh^3 r}\quad
\label{weyl}
\end{equation}
is non-vanishing, while the rest are vanish. Thus the metric is clearly of type D in the Petrov classification scheme.

We set up an orthonormal frame ${\bf e}_{(a)}=\{{\bf e}_{(0)},{\bf e}_{(1)},{\bf e}_{(2)},{\bf e}_{(3)}\}$, ${\bf e}_{(a)}\cdot{\bf e}_{(b)}\equiv e^{\,\mu}_{(a)}\,e^{\,\nu}_{(b)}\,g_{\mu\nu}=\eta_{ab}=\mbox{diag}(-1,+1,+1,+1)$ which consists of three spacelike unit vectors ${\bf e}_{(i)}$, $i=1,2,3$ and one time-like vector ${\bf e}_{(0)}$ \cite{IDS}. Notations are such that small latin indices are raised and lowered with Minkowski metric $\eta^{ab}$, $\eta_{ab}$ and greek indices are raised and lowered with metric tensor $g^{\mu \nu}$, $g_{\mu \nu}$. The dual basis is ${\bf e}^{(i)}={\bf e}_{(i)}$ and ${\bf e}^{(0)}=-{\bf e}_{(0)}$. These frame components in terms of tetrad vector can be expressed as
\begin{equation}
{\bf k}=\frac{1}{\sqrt{2}}\,\left({\bf e}_{(0)}+{\bf e}_{(2)}\right),\quad {\bf l}=\frac{1}{\sqrt{2}}\,\left({\bf e}_{(0)}-{\bf e}_{(2)}\right),\quad {\bf m}=\frac{1}{\sqrt{2}}\,\left({\bf e}_{(1)}+i\,{\bf e}_{(3)}\right).
\label{tetrad}
\end{equation}

In order to analyze the effect of local gravitational fields of these solutions, we have used the equations of geodesic deviation \cite{AOP,EPJC,AHEP2,FAEP1,FAEP2,PS,JP} which in terms of orthonormal frame ${\bf e}_{(a)}$ are
\begin{equation}
\ddot{Z}^{(i)}=-R^{(i)}_{\,(0)(j)(0)}\,Z^{(j)},\quad i,j=1,2,3,
\label{a1}
\end{equation}
where ${\bf e_{(0)}}={\bf u}$ is time-like four-velocity vector of the free test particles. We set here $Z^{(0)}=0$ such that all test particles are synchronized by the proper time. From the standard definition of the Weyl tensor and the Einstein's field equation for zero the stress energy tensor, we get (see Eq. (4) in \cite{JP})
\begin{equation}
R_{(i)(0)(j)(0)}=C_{(i)(0)(j)(0)}-\frac{\Lambda}{3}\,\delta_{ij},
\label{a2}
\end{equation}
where $C_{(i)(0)(j)(0)}\equiv e^{\mu}_{(i)}\,u^{\nu}\,e^{\rho}_{(j)}\,u^{\sigma}\,C_{\mu\nu\rho\sigma}$ are the components of the Weyl tensor.

The only non-vanishing Weyl scalars are given by (\ref{weyl}) so that
\begin{equation}
C_{(1)(0)(1)(0)}=-\Psi_2=C_{(3)(0)(3)(0)}\quad,\quad C_{(2)(0)(2)(0)}=2\,\Psi_2.
\label{a3}
\end{equation}
Therefore, the equations of geodesic deviation (\ref{a1}) takes the following form
\begin{eqnarray}
\ddot{Z}^{(1)}&=&-R^{(1)}_{\,(0)(j)(0)}\,Z^{(j)}=-\left (C_{(1)(0)(1)(0)}-\frac{\Lambda}{3} \right)\,Z^{(1)}=\left (\Psi_2+\frac{\Lambda}{3} \right)\,Z^{(1)},\nonumber\\
\ddot{Z}^{(2)}&=&-R^{(2)}_{\,(0)(j)(0)}\,Z^{(j)}=-\left (C_{(2)(0)(2)(0)}-\frac{\Lambda}{3} \right)\,Z^{(2)}=\left (-2\,\Psi_2+\frac{\Lambda}{3} \right)\,Z^{(2)},\nonumber\\
\ddot{Z}^{(3)}&=&-R^{(3)}_{\,(0)(j)(0)}\,Z^{(j)}=-\left (C_{(3)(0)(3)(0)}-\frac{\Lambda}{3} \right)\,Z^{(3)}=\left (\Psi_2+\frac{\Lambda}{3} \right)\,Z^{(3)}.
\label{a4}
\end{eqnarray}

In the limit $\alpha \rightarrow 0$, all the Weyl scalars including $\Psi_2$ vanishes. In this limit, the space-time (\ref{temp}) becomes anti-de Sitter (AdS) space. So the equations of geodesic deviation (\ref{a4}) in this limit reduces to
\begin{equation}
\ddot{Z}^{(i)}=\frac{\Lambda}{3}\,Z^{(i)}
\label{a6}
\end{equation}
with the solutions
\begin{eqnarray}
&&Z^{(i)}=a_i\,\tau+b_i\quad\quad\quad\quad\quad\quad\quad\quad\quad\quad\quad\quad\quad\quad\, \mbox{for}\quad \Lambda=0,\nonumber\\
&&Z^{(i)}=a_i\,\cos \left(\sqrt{-\frac{\Lambda}{3}}\,\tau \right)+b_i\,\sin \left (\sqrt{-\frac{\Lambda}{3}}\,\tau \right)\quad \mbox{for}\quad \Lambda<0,
\label{a7}
\end{eqnarray}
where $a_i,b_i, i=1,2,3$ are the arbitrary constants.

Again, in the limit $\Lambda \rightarrow 0$, that is, $\beta \rightarrow 0$, the only non-vanishing Weyl scalars is $\Psi_2$ given by (\ref{weyl}). The space-time (\ref{temp}) reduces to type D vacuum space-time of zero cosmological constant with a curvature singularity which we discussed, in detail in Ref. \cite{AHEP}. In this limit ($\Lambda \rightarrow 0$), the equations of geodesic deviation (\ref{a4}) becomes
\begin{equation}
\ddot{Z}^{(1)}=\Psi_2\,Z^{(1)},\quad \ddot{Z}^{(2)}=-2\,\Psi_2\,Z^{(2)},\quad \ddot{Z}^{(3)}=\Psi_2\,Z^{(3)}
\label{a8}
\end{equation}
with the solutions
\begin{eqnarray}
Z^{(1)}&=&c_1\,\cosh \left (\sqrt{\Psi_2}\,\tau \right)+d_1\,\sinh \left (\sqrt{\Psi_2}\,\tau \right),\nonumber\\
Z^{(2)}&=&c_2\,\cos \left(\sqrt{2\,\Psi_2}\,\tau \right )+d_2\,\sin \left (\sqrt{2\,\Psi_2}\,\tau \right),\nonumber\\
Z^{(3)}&=&c_3\,\cosh \left (\sqrt{\Psi_2}\,\tau \right )+d_3\,\sinh \left (\sqrt{\Psi_2}\,\tau \right ),
\label{a9}
\end{eqnarray}
where $c_i, d_i, i=1,2,3$ are the arbitrary constants and $\Psi_2 \neq 0$.

\section{Summary and future work}

In this paper, we generalize a Ricci flat space-time \cite{AHEP} to the case of non-vanishing cosmological constant solution in four-dimensional curved space-time, still represent vacuum solutions of the Einstein's field equations is the generalization of 2D Misner space metric. By introducing a cosmological constant term into the metric components in the metric \cite{AHEP}, we have seen that for $r=r_0$, and $z=z_0$, where $r_0,  z_0$ are constants these space-times reduces to 2D conformal Misner space geometry. As discussed in section {\it 2}, the Misner space metric admit CTCs which appear after a certain instant of time from causally well-behaved conditions. Thus the presented metrics as well as the one studied in \cite{AHEP} evolves CTC from an initial spacelike hypersurface at a certain instant of time. Though causality violating space-times have been studied extensively in literature, few of them belongs to true time-machine space-time ({\it e. g.}, \cite{Ori,EPJP, JPCO,AOP,PTEP2,PTEP,EPJC,CTP2}), and others in ({\it e. g.}, \cite{AHEP,AHEP2,AHEP3,PTEP2}) are lacking one or more basic requirements for a true time-machine space-time. In addition, many time-machine models mentioned in the introduction violate one or more the energy condition. Our space-time is the vacuum solution of the Einstein's field equations of non-zero cosmological constant. So all the energy conditions automatically satisfied and the modified metrics would represent true time-machine space-time but lacking the property of being free from curvature singularity. Furthermore, we have analyzed the space-time and discussed their physical properties. It was demonstrated that these space-times can be understood as gravitational waves of Coulomb-types which propagate on anti-de Sitter backgrounds. A positive cosmological constant ($\Lambda$) plays an importance role in explaining the dynamics of the universe. But in our case, however, it is negative. One can use this modified space-time as a model to study the quantum gravity in connection to the quantum field theory. The dynamic stability of the modified space-times are beyond the scope of this work. Our motivation to further study this problem is to construction a space-time metric which satisfies all criteria for a true time-machine, like obeying the energy conditions, realistic or known types of matter sources, singularity-free and evolves CTCs from an initial spacelike hypersurface in a causally well-behaved manner after a certain instant of time.

\section*{Acknowledgement} Authors sincerely acknowledge the anonymous kind referees for their valuable comments and constructive suggestions.

\section*{Data Availability:} There is no data associated with this manuscript or no data have been used to prepare it.

\section*{Conflict of Interest:} The authors declare that there are no conflict of interest regarding publication  this paper.

\end{document}